\documentclass[12pt]{article}
\usepackage[cp1251]{inputenc}
\usepackage[russian]{babel}
\begin{document}
\large
\par
\begin{center}
{\bf Remarks to the Standard Theory of Neutrino Oscillations.
\par
Corrected Theory of Neutrino Oscillations}  \\
\par
\vspace{0.5cm} Beshtoev Kh. M.
\par
\vspace{0.5cm} Joint Institute for Nuclear Research, Joliot Curie
6, 141980 Dubna, Moscow region, Russia \\
\end{center}
\par
{\bf Abstract} \\

\par
In the Standard theory of neutrino oscillations it is supposed
that physical observed neutrino states $\nu_{e}, \nu_{\mu },
\nu_{\tau}$ have no definite masses and that neutrinos are
initially created as mixture of $\nu_{1}, \nu_{2}, \nu_{3}$
neutrino states and that neutrino oscillations are the real ones
even when neutrino masses are different. It is shown that these
suppositions lead to violation of the law of energy and momentum
conservation and then the neutrino states are unstable ones and
they must disintegrate. Then the development of the standard
theory of neutrino oscillations in the framework of particle
physics is considered where the above mentioned shortcomings are
absent and the oscillations of neutrino with equal masses are real
ones and the oscillations of neutrino different masses are virtual
ones. Expressions for probabilities of neutrino transitions
(oscillations) in the correct theory are given. \\
PACS numbers: 14.60.Pq; 14.60.Lm \\

\par
\section{Introduction}

\par
The suggestion that, in analogy with $K^{o},\bar K^{o}$
oscillations, there could be neutrino-antineutrino oscillations (
$\nu \rightarrow \bar \nu$) was considered by Pontecorvo [1] in
1957. It was subsequently considered by Maki et al. [2] and
Pontecorvo [3] that there could be mixings (and oscillations) of
neutrinos of different flavors (i.e., $\nu _{e} \rightarrow \nu
_{\mu }$ transitions). In the Standard theory of neutrino
oscillations [4] is supposed that physical observed neutrino
states $\nu_{e}, \nu_{\mu }, \nu_{\tau}$ have no definite masses
and that they are directly created as mixture of the $\nu_{1},
\nu_{2}, \nu_{3}$ neutrino states. Below we discuss the
consequences of these suppositions and then the correct
theory of neutrino oscillations is considered.  \\

\par
\section{Remarks to the Theory of Neutrino Oscillations.
Corrected Theory of Neutrino Oscillations}

\par
At first, shortcomings of the Standard theory of neutrino
oscillations are considered and then we pass to consideration of
the corrected theory of neutrino oscillations. \\

\par
\subsection{Remarks to the Standard Theory of Neutrino
Oscillations}

\par
In the Standard theory of neutrino oscillations [4], constructed
in the framework of Quantum theory (Mechanics) in analogy with the
theory of $K^{o}, \bar{K}^{o}$ oscillation, it is supposed that
mass eigenstates are $\nu_{1}, \nu_{2}, \nu_{3}$ neutrino states
but not physical observed neutrino states $\nu_{e}, \nu_{\mu },
\nu_{\tau}$. And that the neutrinos $\nu_{e}, \nu_{\mu },
\nu_{\tau}$ are directly created as superpositions of $\nu_{1},
\nu_{2}, \nu_{3}$ states (neutrinos). Then it is supposed that the
$\nu_{e}, \nu_{\mu }, \nu_{\tau}$ neutrinos have no definite mass,
i.e. their masses may vary in dependence on the $\nu_{1}, \nu_{2},
\nu_{3}$  admixture in the $\nu_{e}, \nu_{\mu }, \nu_{\tau}$
states. And also that neutrino oscillations are real oscillations
even when their masses are different, i.e. that there is a real
transition of electron neutrino $\nu_e$ into muon neutrino
$\nu_{\mu}$ (or tau neutrino $\nu_{\tau}$). Obviously it is
necessary to check up these suppositions. To simplify this, the
case of two neutrinos is considered.
\par
The mass lagrangian of two neutrinos ($\nu_e, \nu_\mu$) has the
following form ($\nu \equiv \nu_L$):
$$
\begin{array}{c}{\cal L}_{M} = - \frac{1}{2} \left[m_{\nu_e}
\bar \nu_e \nu_e + m_{\nu_\mu} \bar \nu_{\mu} \nu_{\mu } +
m_{\nu_e \nu_{\mu }}(\bar \nu_e \nu_{\mu } + \bar \nu_{\mu }
\nu _e) \right] \equiv \\
\equiv  - \frac{1}{2} (\bar \nu_e, \bar \nu_\mu)
\left(\begin{array}{cc} m_{\nu_e} & m_{\nu_e \nu_{\mu }} \\
m_{\nu_{\mu} \nu_e} & m_{\nu_\mu} \end{array} \right)
\left(\begin{array}{c} \nu_e \\ \nu_{\mu } \end{array} \right)
\end{array} ,
\eqno(1)
$$
which is diagonalized by rotation on the angle $\theta$ and then
this lagrangian (1) transforms into the following one (see ref. in
[4]):
$$
{\cal L}_{M} = - \frac{1}{2} \left[ m_{1} \bar \nu_{1} \nu_{1} +
m_{2} \bar \nu_{2} \nu_{2} \right]  , \eqno(2)
$$
where
$$
m_{1, 2} = {1\over 2} \left[ (m_{\nu_e} + m_{\nu_\mu}) \pm
\left((m_{\nu_e} - m_{\nu_\mu})^2 + 4 m^{2}_{\nu_\mu \nu_e}
\right)^{1/2} \right] ,
$$
\par
\noindent and angle $\theta $ is determined by the following
expression:
$$
tg 2 \theta  = \frac{2 m_{\nu_e \nu_\mu}} {(m_{\nu_\mu} -
m_{\nu_e})} , \eqno(3)
$$
$$
\begin{array}{c}
\nu_e = cos \theta  \nu_{1} + sin \theta \nu_{2}  ,         \\
\nu _{\mu } = - sin \theta  \nu_{1} + cos \theta  \nu_{2} .
\end{array}
\eqno(4)
$$
Then $\nu_e, \nu_\mu$ masses are:
$$
m_{\nu_e} = m_1 cos^2 \theta + m_2 sin^2 \theta ,
$$
$$
m_{\nu_\mu} = m_1 sin^2 \theta + m_2 cos^2 \theta , \eqno(5)
$$
i.e., $\nu_e, \nu_\mu$ neutrinos have definite masses which are
expressed via $\nu_1, \nu_2$ masses and mixing angle $\theta$ . It
means that supposition that $\nu_e, \nu_\mu$ neutrinos have no
definite masses is not confirmed.
\par
If neutrino oscillations are real oscillations, i.e. there is a
real transition of electron neutrino $\nu_e$ into muon neutrino
$\nu_{\mu}$ (or tau-$\nu_{\tau}$ neutrino). Then the neutrino $x =
\mu, \tau$ will decay in electron neutrino plus something
$$
\nu_{x} \rightarrow \nu_e + ....  , \eqno(6)
$$
as a result, we get energy from vacuum, which is equal to the mass
difference (if $m_{\nu_x} > m_{\nu_e}$)
$$
\Delta E \sim m_{\nu_{x}} - m_{\nu_e} . \eqno(7)
$$
Then, again this electron neutrino transits into the muon
neutrino, which decays again and we get energy and etc. {\bf So we
have got a perpetuum mobile!} Obviously, {\bf the law of energy
and momentum conservation} in these processes is not fulfilled.
\par
Besides, since $\nu_e, \nu_{\mu}, \nu_{\tau}$ neutrinos are
superpositions of $\nu_{1}, \nu_{2}, \nu_{3}$ then the $\nu_e,
\nu_{\mu}, \nu_{\tau}$ neutrinos are wave packets having widths
which are equal to mass differences of the composing components,
i.e., $ \nu_{1}, \nu_{2}, \nu_{3}$ neutrinos, $\Delta m \to
m_{\nu_2} - m_{\nu_1}$ or $m_{\nu_3} - m_{\nu_1}$ . Then these
$\nu_e, \nu_{\mu}, \nu_{\tau}$ states (neutrinos) are unstable
ones and must disintegrate for the time $t$ which is determinated
by the uncertainty relation [5], i.e.,
$$
t \cong \frac{1}{\Delta m} . \eqno(8)
$$
It also means that the Sun neutrinos cannot reach the Earth as
$\nu_e, \nu_{\mu}, \nu_{\tau}$ neutrino states. But in experiments
[6, 7] we see namely $\nu_e, \nu_\mu, \nu_\tau$ states (neutrinos)
but not other states. Without any doubt this Standard theory
requires a correction in order to get rid of the above mentioned
defects. \\

\par
\subsection{Corrected Theory of Neutrino Oscillations}
\par

In the framework of the Quantum Mechanics [5] all the states are
wave packets having widths and these states are unstable ones and
they must disintegrate. In contrast to the Quantum Mechanics in
the framework of the  Particle Physics theory [8] all particles
are stable ones or if they have widths then they must decay in the
states (particles) with small masses.
\par
The only way to restore the law of energy conservation is to
demand that this process is virtual one if neutrinos have
different masses. Then, these oscillations will be virtual ones
and they are described in the framework of the uncertainty
relations.
\par
So, the correct theory of neutrino oscillations can be constructed
only into the framework of the Particle Physics theory, where the
conception of mass shell is present [8, 9]. Besides, every
particle must be created on its mass shell and it will be left on
its mass shell while passing through vacuum.
\par
In the considered theory of neutrino oscillations [9], constructed
in the framework of the particle physics theory, it is supposed
(following to the experiment) that:
\par
1)  The  physical  observable neutrino states $\nu_{e}, \nu_{\mu
}, \nu_{\tau}$ are eigenstates of the weak interaction with $W,
Z^o$ exchanges. And, naturally, the mass matrix of $\nu_{e},
\nu_{\mu }, \nu_{\tau}$ neutrinos is diagonal, i.e., the mass
matrix of $\nu_e, \nu_\mu$ and $\nu_\mu$ neutrinos has the
following diagonal form (since these neutrinos are created in the
weak interactions it means that they are eigenstates of these
interactions and their mass matrix must be diagonal):
$$
\left(\begin{array}{ccc} m_{\nu_e}& 0& 0 \\ 0 & m_{\nu_\mu} & 0\\
0 & 0 & m_{\nu_\mu} \end{array} \right) . \eqno(9)
$$
Besides, all  the  available, experimental results indicate that
the lepton numbers $l_{e}, l_{\mu }, l_{\tau}$  are   well
conserved, i.e. the standard weak interactions (with $W, Z^o$
bosons) do not violate the lepton numbers.
\par
2) Then, to violate the  lepton  numbers, it  is  necessary  to
introduce an interaction violating these numbers. It is equivalent
to introducing of the nondiagonal  mass terms  in the  mass matrix
of $\nu_{e}, \nu_{\mu }, \nu_{\tau}$ neutrinos:
$$
M(\nu_e, \nu_\mu, \nu_\tau) = \left(\begin{array}{ccc} m_{\nu_e}&
m_{\nu_e \nu_\mu} & m_{\nu_e \nu_\tau}
\\ m_{\nu_\mu \nu_e} & m_{\nu_\mu} & m_{\nu_\mu \nu_\tau} \\
m_{\nu_\tau \nu_e} & m_{\nu_\tau \nu_\mu} & m_{\nu_\mu}
\end{array} \right) . \eqno(10)
$$
Diagonalizing this matrix [4]
$$
M(\nu_e, \nu_\mu, \nu_\tau) = V^{-1} M(\nu_1, \nu_2, \nu_2) V ,
\eqno(11)
$$
we go to the $\nu_{1}, \nu _{2}, \nu_{3}$ neutrino mass matrix
$$
\left(\begin{array}{ccc} m_{\nu_1}& 0& 0 \\ 0 & m_{\nu_2} & 0\\
0 & 0 & m_{\nu_3} \end{array} \right) , \eqno(12)
$$
where $V$ is neutrino mixings matrix $V$. Then the vector state
$\Psi(\nu_e, \nu_\mu, \nu_\tau)$, of $\nu_e, \nu_\mu, \nu_\tau$
neutrinos
$$
\Psi(\nu_e, \nu_\mu, \nu_\tau) = \left(\begin{array}{c} \nu_e \\
\nu_\mu \\ \nu_\tau
\end{array} \right) , \eqno(13)
$$
is transformed in the vector state $\Psi(\nu_1, \nu_2, \nu_2)$ of
$\nu_1, \nu_2, \nu_2$ neutrinos
$$
\Psi(\nu_e, \nu_\mu, \nu_\tau) = V \Psi(\nu_1, \nu_2, \nu_2) .
\eqno(14)
$$
In the parameterization proposed by Maiani [10] $V$ has the
following form:
$$
{V = \left( \begin{array} {ccc}1& 0 & 0 \\
0 & c_{\gamma} & s_{\gamma} \\ 0 & -s_{\gamma} & c_{\gamma} \\
\end{array} \right) \left( \begin{array}{ccc} c_{\beta} & 0 &
s_{\beta} \exp(-i\delta) \\ 0 & 1 & 0 \\ -s_{\beta} \exp(i\delta)
& 0 & c_{\beta} \end{array} \right) \left( \begin{array}{ccc}
c_{\theta} & s_{\theta} & 0 \\ -s_{\theta} & c_{\theta} & 0 \\ 0 &
0 & 1 \end{array}\right)} , \eqno(15)
$$
where $\theta, \beta, \gamma$ and $\delta$ are angles of neutrino
mixings and parameter of $CP$ violation.
\par
Exactly like the case  of $K^{o}$ mesons created  in strong
interactions, when mainly $K^{o}, \bar{K}^{o}$ mesons are produced
but not $K_1, K_2$ mesons. In  the considered case $\nu_{e},
\nu_{\mu }, \nu_{\tau}$, but not $\nu_{1}, \nu_{2}, \nu_{3}$,
neutrino  states are mainly created in the weak interactions (this
is so since contribution of the lepton numbers violating
interactions  in this process is too small and in this case no
oscillations take place).
\par
3) Then, when the $\nu_{e}, \nu_{\mu }, \nu_{\tau}$  neutrinos are
passing through vacuum, they  will  be  converted  into
superpositions  of  the $\nu_{1}, \nu _{2}, \nu_{3}$  owing  to
the presence  of  the interactions violating  the  lepton number
of neutrinos and  will be left on  their mass   shells.  And,
then, oscillations of the $\nu_{e}, \nu_{\mu}, \nu_{\tau}$
neutrinos will  take  place according to the standard scheme [4].
Whether these oscillations are real or virtual, it will be
determined by the masses of the  physical observed neutrinos
$\nu_{e}, \nu_{\mu}, \nu_{\tau}$.
\par
i) If the masses of the $\nu_{e}, \nu_{\mu }, \nu_{\tau}$
neutrinos  are equal, then the real oscillation of the neutrinos
will take  place.
\par
ii) If  the masses  of  the $\nu_{e}, \nu _{\mu }, \nu _{\tau}$
are  not equal, then the virtual oscillation of  the  neutrinos
will  take place. To make these oscillations  real,  these
neutrinos must participate  in the quasielastic interactions, in
order to undergo transition  to  the mass shell of the other
appropriate neutrinos in analogy with $\gamma  - \rho ^{o}$
transition  in the  vector   meson  dominance model. It is
necessary to take into account that in contrast to the strong
interactions, the dependence on squared transferring momentum in
the weak interactions has flat form since $W$ boson has a huge
mass. In case ii) enhancement of neutrino oscillations will take
place if the mixing angle is small at neutrinos passing through  a
bulk of matter [11].
\par
So the neutrino mixings (oscillations) appear due to the fact that
at neutrino creating the eigenstates of the weak interactions
(i.e. $\nu_e, \nu_\mu, \nu_\tau$ neutrino states) are produced
 but not the eigenstates of the weak interaction violating
lepton numbers (i.e. $\nu_1, \nu_2, \nu_3$ neutrino states). And
then when neutrinos are passing through vacuum they are converted
into superpositions of $\nu_1, \nu_2, \nu_3$ neutrinos. If $\nu_1,
\nu_2, \nu_3$ neutrinos were originally created, then the mixings
(oscillations) would not have taken place since the weak
interaction conserves the lepton numbers.
\par
In the case of three neutrino types the probability of $\nu_e \to
\nu_e$ transitions  has the following form:
$$
P(\nu_e \to \nu_e, t)= 1 - cos^4(\beta)sin^2(2 \theta) sin^2(-t
(E_1-E_2)/2) -
$$
$$
cos^2(\theta) sin^2(2 \beta) sin^2(-t (E_1-E_3)/2) -
$$
$$
- sin^2(\theta) sin^2(2 \beta) sin^2(-t (E_2-E_3)/2) , \eqno(16)
$$
where $E_1, E_2, E_3$ are energy of $\nu_1, \nu_2, \nu_3 \to x$
neutrinos and $E_x = \sqrt{p^2 + m^2_x}$.
\par
Since lengths of neutrino oscillations
$$
L_{i, j} = 2\pi  {2p \over {\mid m^{2}_{2} - m^{2}_{1} \mid}}
\quad i \ne j = 1, 2, 3 . \eqno(17)
$$
are different, then the expression of probability for neutrino
oscillations at small distances has a simpler form. For example,
for $\nu_e \to \nu_e$ oscillations we have
\par
$$
P(\nu_e \rightarrow \nu_e) = 1 -  \sin^{2} 2\theta sin^2
((m^{2}_{2} - m^{2}_{1}) / 2p) t , \eqno(18)
$$
where
$$
sin^2 \theta= 1/2 - \frac{(m_{\nu_e} - m_{\nu_\mu})}{2
\sqrt{(m_{\nu_e} - m_{\nu_\mu})^2 +(2 m_{\nu_e \nu_\mu})^2}} ,
\eqno(19)
$$
and

$$
sin^2 2\theta = \frac{(2m_{\nu_{e} \nu_{\mu}})^2} {(m_{\nu_e} -
m_{\nu_\mu})^2 +(2m_{\nu_e \nu_{\mu}})^2} , \eqno(20)
$$
\par
It is interesting to remark that expression (20) can be obtained
from the Breit-Wigner distribution [12]
$$
P \sim \frac{(\Gamma/2)^2}{(E - E_0)^2 + (\Gamma/2)^2}   ,
\eqno(21)
$$
by using the following substitutions:
$$
E = m_{\nu_e},\hspace{0.2cm} E_0 = m_{\nu_\mu},\hspace{0.2cm}
\Gamma/2 = 2m_{\nu_e, \nu_\mu} , \eqno(22)
$$
where $\Gamma/2 \equiv W(... )$ is a width of $\nu_e
\leftrightarrow \nu_\mu$ transitions, i.e., virtual neutrino
oscillations keeps in within the uncertainty relation. In the
general case these widths can be computed by using a standard
method [13].
\par
If $m_{\nu_e, \nu_\mu}$ differs from zero, then Exp. (20) gives a
probability of $\nu_e \leftrightarrow \nu_\mu$ transitions and
then the probability of $\nu_e \leftrightarrow \nu_\mu$
transitions is defined by these neutrino masses and width of their
transitions. If $m_{\nu_e, \nu_\mu} = 0$, then the ${\nu_e
\leftrightarrow \nu_\mu}$ transitions are forbidden. So, this is a
solution of the problem of the origin of the mixing angle in the
theory of vacuum oscillations.
\par
It is necessary to remark that in this correct theory of neutrino
oscillations, in contrast to the Standard theory, oscillations of
neutrinos with equal masses are real ones and the oscillations of
neutrinos with different masses are virtual ones and then the
problem of energy momentum conservation and the problem of
neutrino disintegrations as wave packets are solved.
\par
In the above considered theory of neutrino oscillations neutrino
masses change at neutrino oscillations (for example $m_{\nu_e} \to
m_{\nu_\mu}$). Theoretically it is also possible neutrino
transitions without changing their masses [13]. In this case the
mixing angles are maximal ($\pi/4$). The author proposed another
mechanism (model) of neutrino transitions which is analogous to
the model of vector dominance, i.e., the model of $\gamma \to
\rho^o$ transitions [14]. \\

\par
\section{Conclusions}

\par
In the Standard theory of neutrino oscillations it is supposed
that physical observed neutrino states $\nu_{e}, \nu_{\mu },
\nu_{\tau}$ have no definite masses and that they are initially
created as mixture of the $\nu_{1}, \nu_{2}, \nu_{3}$ neutrino
states and that neutrino oscillations are real ones even when
their masses are different. It was shown that these suppositions
lead to violation of the law of energy and momentum conservation
and then the neutrino states are unstable ones and they must
disintegrate. Then the development of the Standard theory of
neutrino oscillations in the framework of particle physics has
been considered where the above mentioned shortcomings are absent
and the oscillations of neutrino with equal masses are real ones
and the oscillations of neutrino different masses are virtual
ones. Expressions for probabilities of neutrino transitions
(oscillations) in the corrected theory
have been given. \\

\par
{\bf References}\\

\par
\noindent 1. Pontecorvo B. M., Soviet Journ. JETP, 1957, v. 33,
p.549;
\par
JETP, 1958,  v.34, p.247.
\par
\noindent 2. Maki Z. et al., Prog.Theor. Phys., 1962, vol.28,
p.870.
\par
\noindent 3. Pontecorvo B. M., Soviet Journ. JETP, 1967, v. 53,
p.1717.
\par
\noindent 4. Bilenky S.M., Pontecorvo B.M., Phys. Rep.,
C41(1978)225;
\par
Boehm F., Vogel P., Physics of Massive Neutrinos: Cambridge
\par
Univ. Press, 1987, p.27, p.121;
\par
Bilenky S.M., Petcov S.T., Rev. of Mod.  Phys., 1977, v.59,
\par
p.631.
\par
 Gribov V., Pontecorvo B.M., Phys. Lett. B, 1969, vol.28,
p.493.
\par
\noindent 5. Schiff L. I., Quantum Mechanics, McRam, ..., London,
1955.
\par
Kayser B., Phys. Rev. D24, 1981, p.110.
\par
\noindent 6. Kameda J., Proceedings of ICRC 2001, August 2001,
Germany,
\par
Hamburg, p.1057.
\par
Fukuda  S. et al,. Phys.   Rev. Lett., 2001, v.25, p.5651;
\par
Phys. Lett. B, 539, 2002,  p.179.
\par
\noindent 7. Ahmad Q. R. et al., Internet Pub. nucl-ex/0106015,
June 2001.
\par
Ahmad  Q. R. et al., Phys. Rev. Lett. 2002, v. 89, p.011301-1;
\par
Phys. Rev. Lett.  2002,v.  89, p.011302-1.
\par
\noindent
8. Schweber S., An Introduction to Relativistic Quantum
 Field Theory,
\par
 Row, ..., New York, 1961.
\par
 Ta-Pei Cheng, Ling-Fong Li, Gauge Theory of Elementary
Particle
\par Physics, Clarendon Press-Oxford, 1984.
\par
\noindent 9. Beshtoev Kh.M., JINR Commun. E2-92-318, Dubna, 1992;
\par
JINR Rapid Communications, N3[71]-95.
\par
HEP-PH/9911513;
\par
 The Hadronic Journal, v.23, 2000, p.477;
\par
Proceedings of 27th Intern. Cosmic Ray Conf., Germany,
\par
Hamburg, 7-15 August 2001, v.3, p. 1186.
\par
\noindent 10. L. Maiani, Proc. Int. Symp.  on  Lepton-Photon
\par
Inter., Hamburg, DESY, p.867.
\par
\noindent 11. Beshtoev Kh.M., JINR Commun, E2-93-297, Dubna, 1993;
\par
JINR Commun. E2-94-46; Hadronic Journal, 1995, vol 18, p.165.
\par
\noindent 12. Blatt J.M., Weiscopff V.F., The Theory of Nuclear
Reactions,
\par
INR T.R. 42.
\par
\noindent 13. Beshtoev Kh.M., HEP-PH/9911513;
\par
 The Hadronic Journal, v.23, 2000, p.477;
\par
Proceedings of 27th Intern. Cosmic Ray Conf., Germany,
\par
Hamburg, 7-15 August 2001, v.3, p. 1186.
\par
Beshtoev Kh.M., JINR Commun. E2-99-307, Dubna, 1999;
\par
JINR Commun. E2-99-306, Dubna, 1999.
\par
\noindent 14. Sakurai  J.J., Currents  and  Mesons,  The  Univ of
Chicago Press, 1967.
\par
Beshtoev Kh. M., JNR of USSR Academy Science P-217,
\par
Moscow, 1981.

\end{document}